# Stochastic Theory of Dust-Grain Charging in Low-Pressure Plasmas

S.N. Abolmasov, E.V. Romashchenko, and P. Roca i Cabarrocas
*LPICM, CNRS (UMR 7647), Ecole Polytechnique, 91128 Palaiseau, France*

Charging of dust grains in low-pressure plasmas is reviewed critically. A theory based on the Fokker-Planck equation and orbital motion limited approximation is proposed. The theory predicts that dust grains can acquire a positive charge in low-pressure electropositive plasmas having a sufficiently high plasma potential, in agreement with experimental observations. It is also shown that variations in the plasma potential (electron temperature) can lead to spatial regions in which grains have opposite charges.

The presence of dust grains is quite common in both space and laboratory plasmas. It is then not surprising that plasmas containing dust particles have been the subject of intensive studies over the past five decades and still attract attention of many researchers. The original interest in understanding the properties of dusty plasmas was essentially driven by aerosol and space plasma science communities.[1-3] The discovery of dust crystals in 1994 resulted in a surge of interest in complex plasmas.[4,5] Current interest derives from the control of the dust particle growth and transport in various plasma processing and fusion reactors.[6-12]

The particle charge is one of the most important parameters of dusty plasmas. It determines the particle interactions with background plasma and neighbour particles. Therefore, a calculation of the charge on a particle is the starting point of every theory of dusty plasmas. The charge arises from collecting of electrons and ions from the plasma and sometimes from emitting electrons (e.g., due to secondary, thermionic or photoelectric emission). The charging currents to a particle depend on the potential difference between particle and plasma rather than charge. Therefore it is usually more convenient to discuss the charge on a particle in terms of the corresponding potential with respect to the surrounding medium- plasma. The orbital motion limited (OML) theory, originally developed for electrostatic probes,[13] is generally used to find the potential of dust particles immersed in plasma.[14] In this case, dust particles act as a small spherical probe at floating potential $\varphi_f$, at which the probe draws no current.

It is generally believed that in a plasma in which emission processes are unimportant, the equilibrium charge on a dust grain, and its surface potential $\varphi_d$, is negative because the flux of electrons to a floating surface is high relative to that of ions. On the other hand, when electron emission is significant, the equilibrium charge can become positive, as shown by OML models and well documented in reviews on the subject.[9-12] In this Letter, a theory is outlined predicting that this is only true in space plasmas, whereas in some laboratory plasmas dust particles can be positively charged even in the absence of electron emission.

*Stochastic theory.–* Electrons and ions are absorbed at the particle surface at random times, casing the surface charge to fluctuate around an equilibrium value $Q_0$.[15] Since in plasmas the ions are usually singly charged, the charge fluctuations can be described as a one-step stochastic process, which is a subclass of the Markov processes.[16] This approach is particularly relevant to small grains (< 1 μm) near which a fluid (i.e., Vlasov) description of the plasma breaks down due to the inherent discreteness of the plasma.[17,18] Following the work of Matsoukas and Russell,[19,20] a linear Fokker-Planck equation is used to describe the charge fluctuations:

$$\frac{\partial P(Z,t)}{\partial t} = -\frac{\partial}{\partial Z} A(Z)P + \frac{1}{2} B \frac{\partial^2 P}{\partial Z^2}, \tag{1}$$

where $P(Z, t)$ is the probability per unit time for a particle to carry charge $Q = eZ$ (with $Z = \pm 1, 2, \ldots$). In our case, we consider an ensemble of identical spherical particles immersed in an electropositive plasma. The particles have a geometric radius $a$, which is smaller than the Debye length $r_D$ – the nominal radius of the shielding cloud around the particle ($a \ll r_D$). The dust-particle concentration $n_d$ is assumed to be sufficiently low (typically $n_d < 10^5$ cm$^{-3}$) that an inter-particle distance $d = n_d^{-1/3}$ is larger than the Debye length ($r_D \sim 10\text{-}100$ μm in laboratory plasmas). In other words, the particles are completely screened by their surrounding shielding clouds and the particle-particle interactions which reduce the charge on the particle[21] can be ignored. The Debye length is also assumed to be less than a mean free path for ion and electron collisions ($r_D < \lambda$). Consequently, the coefficients $A$ and $B$ can be derived in the framework of the OML approximation (generalization to other charging mechanisms is trivial).

Matsoukas and Russell[19] originally derived an equation similar to Eq. (1) by expanding the so-called master equation which was used to model charging of dust particles as a one-step stochastic process. In doing so, they arrived at the equation in which the coefficient $B$ [$C^2/s$] had no physical meaning; the same is true for a recent work of Shotorban.[22] Nevertheless, in a later work, Matsoukas and Russell[20] derived a linear Fokker-Planck equation in the correct form through the linearization of the charging currents in the vicinity of the steady-state charge $Q_0$. This has been achieved through introducing a characteristic time scale, which they called the fluctuation time $\tau_f$. Surprisingly, $\tau_f$ was a strong function of the ion parameters, but nearly independent of the electron temperature. In the present work, we will use a different (phenomenological) approach and show that the Fokker-Planck equation can provide a deeper insight into the physics of the process. In particular, the plasma potential $\varphi_{pl}$ explicitly enters into the Fokker-Planck equation, as explained below.

Notice that Eq. (1) is equivalent to the classical diffusion equation with drift. Consequently, the charge of dust particle can be viewed as a stochastic variable that exhibits a drift in the $Z$-space superimposed on the random diffusion (fluctuations). The drift term in Eq. (1), involving $A(Z)$, is defined by the deterministic charging currents with $A(Z) = I_i - I_e$, where $I_e$ and $I_i$ are the electron and ion current to the dust particle, respectively. For the collection of Maxwellian electrons and ions, characterized by temperatures $T_e$ and $T_i$, the OML theory implies:[23]

$$I_e = \pi a^2 e n_0 \left(\frac{8kT_e}{\pi m_e}\right)^{1/2} \exp\left(\frac{e(\varphi_d - \varphi_{pl})}{kT_e}\right) \tag{2}$$

$$I_i = \pi a^2 e n_0 \left(\frac{8kT_i}{\pi m_i}\right)^{1/2} \left(1 - \frac{e(\varphi_d - \varphi_{pl})}{kT_i}\right), \tag{3}$$

where $n_0$ is the plasma density, $k$ is the Boltzmann constant, and $m_{e,i}$ is the electron (ion) mass. Note that in plasmas, in which emission processes are unimportant, the floating potential (or equivalently the particle's surface potential) is always negative with respect to the plasma potential, i.e. $\varphi_d - \varphi_{pl} < 0$. It should also be mentioned that it is often the case in literature that the authors do not explicitly write the potential difference between the dust particle and the plasma, preferring instead the short-hand: $\varphi_d - \varphi_{pl} \equiv \varphi_d = \varphi_f$. One must always remember, however, that in this definition $\varphi_d$ or $\varphi_f$ is the surface potential of the particle relative to the plasma potential, which is *not necessarily equal to zero*.

It should be emphasized that in practice all potentials are measured with respect to the reference potential (ground in the laboratory). Therefore, in the laboratory framework the potential difference is defined by $\varphi_d - \varphi_{pl} = [\varphi_d - \varphi(0)] - [\varphi_{pl} - \varphi(0)]$, where $\varphi(0)$ is the ground potential. The charge $Q = eZ$ is then related to the particle's surface potential $\varphi_d$ by

$$Q = C_d \varphi_d = 4\pi\varepsilon_0 a \varphi_d, \tag{4}$$

where $C_d$ is the *self*-capacitance of the particle and $\varepsilon_0$ is the vacuum permittivity. It should also be stressed, that it is a common mistake to define $Q$ by using the capacitance of two concentric spheres[15,23,24]

$$Q = C(\varphi_d - \varphi_{pl}) = \frac{4\pi\varepsilon_0 a}{1 - a/r_D}(\varphi_d - \varphi_{pl}). \tag{5}$$

This is simply because this equation describes *induced* charges, not the charge caused by collection of electrons and ions. Note that Eq. (5) was originally proposed by Whipple[25] to describe the charge of dust grains in space, where $\varphi_{pl} = 0$ and in the small particle limit $a/r_D \ll 1$ Eq. (5) finally reduces to Eq. (4).

It is possible to show that if $A(Z) < 0$, then the stationary solution of Eq. (1) is Gaussian given by[16]

$$P_s(Z) = \frac{const}{B} \exp\left[\frac{2}{B}\int_0^Z A(Z')dZ'\right], \tag{6}$$

with the integration constant defined by

$$\int_{-\infty}^{+\infty} P_s(Z')dZ' = 1. \tag{7}$$

Inserting Eqs. (2)-(4) into Eq. (1) with $\partial P/\partial t = 0$ and introducing the following dimensionless parameters

$$\alpha = \left(\frac{m_i}{m_e}\frac{T_e}{T_i}\right)^{1/2}, \quad \beta = \frac{T_e}{T_i}, \quad \psi = \frac{e^2 Z}{4\pi\varepsilon_0 akT_e}, \quad \text{and } \hat{\psi} = \frac{e\varphi_{pl}}{kT_e} \tag{8}$$

yield the final Fokker-Planck equation

$$-\frac{\partial}{\partial \psi}[1 - \beta(\psi - \hat{\psi}) - \alpha\exp(\psi - \hat{\psi})]P + \frac{1}{2}\frac{\partial^2 P}{\partial \psi^2} = 0, \tag{9}$$

with the initial diffusion coefficient $B$ defined as

$$B = \left(\frac{4\pi\varepsilon_0 akT_e}{e}\right)^2 \bigg/ \tau. \tag{10}$$

In this formulation $\tau$ represents a linear charge relaxation time given by

$$\tau = K \frac{\sqrt{T_e}}{a n_0}, \qquad (11)$$

where $K$ is a function of $T_e/T_i$ and $m_i$. The dependence of the relaxation time on the grain size can be understood from the capacitance model[24,25] with $C \propto a$ (Eq. (4)) and $R \propto a^{-2}$, which makes $\tau = RC \propto a^{-1}$. The resistor $R$ here is related to the slope of the characteristic of a spherical probe at the floating potential. Similar dependence can also be found in the work of Cui and Goree.[15]

*Results and discussion.–* The charge distribution functions calculated using Eq. (9) for a hydrogen plasma with parameters typical for the Earth's ionosphere are depicted in Fig. 1(a) for 15, 50, and 150 nm particles. The charge distributions have peaks that are centered near the average charge <Z>. It can be seen that <Z> corresponds to $\varphi_d = -2.5kT_e/e$ which is the well-known Spitzer potential.[3] Note that $\varphi_d$ is independent of the dust particle's size, as predicted by the OML theory. On the other hand, the charge distribution is wider for larger $a$, as also shown by other researchers,[15,19,20] and their height is an inverse function of $a$ according to Eqs. (6), (10) and (11). In the laboratory, a plasma with similar parameters, i.e. $T_e = T_i = 0.2$ eV and $\varphi_{pl} \approx 0$, can be created in so-called Q-machines using alkali metals.[26] For comparison, Fig. 1(b) shows the charge distribution functions predicted for 15, 50, and 150 nm particles immersed in a potassium plasma. In this case, <Z> is more negative due to larger mass of potassium ions while $\varphi_d = -4kT_e/e$. Fig. 2 (curve 1) illustrates a typical probe characteristic of such a plasma, indicating that $\varphi_{pl} \approx 0$ and $\varphi_f < 0$.

It should be emphasized, however, that unlike space plasmas where there is no a reference electrode, plasma in the laboratory is often in contact with grounded surface(s), as shown schematically in the inserts in Figs. 1(a) and 3(a). Consequently, in space one can assume that $\varphi_{pl} = 0$ whereas in the laboratory the magnitude of space potential (determined by a balance of electron and ion creation and loss) depends on the plasma production method. Most laboratory plasmas have a positive (relative to ground) plasma potential that increases in the following order: Q-machine (~ 0 V)[26] < ECR/ICP discharge (10-30 V)[27,28] < CCP/dc glow discharge (> 20 V),[29] where dc, ECR, ICP, and CCP stand for direct current, electron cyclotron resonance, inductively and capacitively coupled plasma, respectively.[30] Since $\varphi_f$ is few $kT_e/e$ lower than $\varphi_{pl}$, one would expect $\varphi_f > 0$ in some laboratory plasmas, particularly in CCP discharges and dc glow discharges with a dc potential applied to the anode. This can also be seen in Fig. 2, where the second curve shows an *I-V* curve of an active Langmuir probe in an argon CCP with $\varphi_f > 0$. In addition, positively charged dust particles have also been observed experimentally, e.g., in the anode region of an abnormal glow discharge in air.[32] Using an analytical model the authors concluded that the positive charge was mainly caused by photoelectric emission. However, the magnitude of UV photon flux used in the model (> 30 mW/cm$^2$) was obviously unrealistic. Such high radiation fluxes can be achieved in high-density plasmas of noble gases, like argon ICPs,[33] in which more than 40% of the electron energy goes into UV production. On contrary, in a dc discharge in air, which mostly consists of two molecular gases (78% $N_2$ and 21% $O_2$), the electron energy is mainly spent (in inelastic collisions) on nitrogen gas heating while UV photons are easily quenched by oxygen.[34]

Fig. 3(a) shows the charge distribution functions predicted by Eq. (9) for 10 nm particles immersed in an argon plasma with parameters ($T_e = 4$ eV, $\varphi_{pl} \geq 20$ V) typical for CCP discharges in the α-regime[29,34] and ICP discharges at low pressures (~ 10 mTorr).[27,28] In this case $\hat{\psi} \neq 0$ and the particle's surface potential $\varphi_d$ is indeed positive. At $\varphi_{pl} = 20$ V the positive charge on a 10 nm particle is about 72 elementary charges and it increases with increasing the plasma potential. Nevertheless, the potential difference $\Delta\varphi = \varphi_{pl} - \varphi_d$ is always a constant determined by the electron temperature; referring to Fig. 3(a), $\Delta\varphi = 2.4kT_e/e$. Note that similar dependence $\Delta\varphi = \varphi_{pl} - \varphi_f \approx 4.7kT_e/e$ follows from a planar probe theory for an argon plasma with Maxwellian electrons.[29] The question arises about how the particle can be positively charged in plasmas in which the electron mobility is higher than the ion mobility. The answer to this question is the net (positive) space charge in the near-electrode sheath (e.g., the cathode sheath in dc discharges). Note that the net space charge actually creates the plasma potential. Since the plasma behaves like a dielectric medium[30,34] the net space charge will induce negative surface charges on the particle, resulting in stronger ion acceleration and electron deceleration which, in turn, can result in a positive collected charge. Consequently, it is the magnitude of plasma potential (as well as electron temperature) that determines the polarity of particle charge rather than the electron mobility. Nevertheless, the net particle charge- the sum of induced and collected charges- will always be negative. Therefore, in order to determine the charge collected by the particle surface Eq. 4 should be used rather than Eq. 5. On the other hand, the electron temperature and plasma potential can also vary in space and time. For example, in argon ICP discharges at 10 mTorr the electron temperature and plasma potential both decrease in the radial direction, lowering to $T_e = 3$ eV and $\varphi_{pl} = 6$ V at the electrode periphery.[27,35] At such plasma parameters particles become negatively charged, as shown in Fig. 3(b), i.e. the discharge may contain spatial regions in which particles have opposite charges. This in fact can explain "abnormal" particle coagulation in the plasma that proceeds at a higher rate than that predicted by Brownian motion- an issue still being actively debated.[36,37]

*Limitations.* − There are circumstances when the OML theory as well as Eq. (9) is not applicable. The first limitation which has been ignored by many authors is due to $d > 2r_D$. This condition is generally fulfilled in electropositive plasmas in which dust grains are externally introduced into the plasma (using a sputtering or a dispenser electrode) and the number of grains is relatively small. In electronegative plasmas, however, particles can grow inside the plasma and their size and number density can vary greatly over time and space. For example, experimental studies by different research groups[6,7,10] conducted in the 1990s revealed that the temporal evolution of particles in silane-containing CCPs is a multistep process. It begins with a brief nucleation phase during which initial crystallites grow within the plasma up to a critical number density (typically $10^9$-$10^{10}$ cm$^{-3}$). These primary particles are rather small (2-20 nm) and monodisperse in size. Once the critical density is reached, a phase of rapid particle growth by coagulation of primary particles sets in. During the coagulation phase the particles grow to a size of about 50-200 nm while their density drastically decreases (below $10^8$ cm$^{-3}$). Further agglomeration is believed to be terminated by particle charging and particles continue to grow by molecular sticking of SiH$_x$ clusters. Obviously, at such high densities of particles the Debye screening is absent and hence the OML is not valid despite the fact that $a << r_D < \lambda$ could be fulfilled. The second limitation derives from the fact that Eq. (9) is only valid when the stochastic process under consideration is slow,[16,38] i.e. at $4\pi\varepsilon_0 akT_e/e^2 >> 1$. Fig. 4 shows that this condition is violated for small particles (below 10 nm in size), resulting in $P > 1$. Therefore, deeper understanding of physical processes in plasmas containing a large amount of small particles will require new advances in both modelling and diagnostics.[39,40]

The authors are grateful to André Bouchoule and Laifa Boufendi for helpful discussions. E.V.R. thanks the Ministry of Education and Science, Youth and Sport of Ukraine for financial support during her stay at LPICM, Ecole Polytechnique.

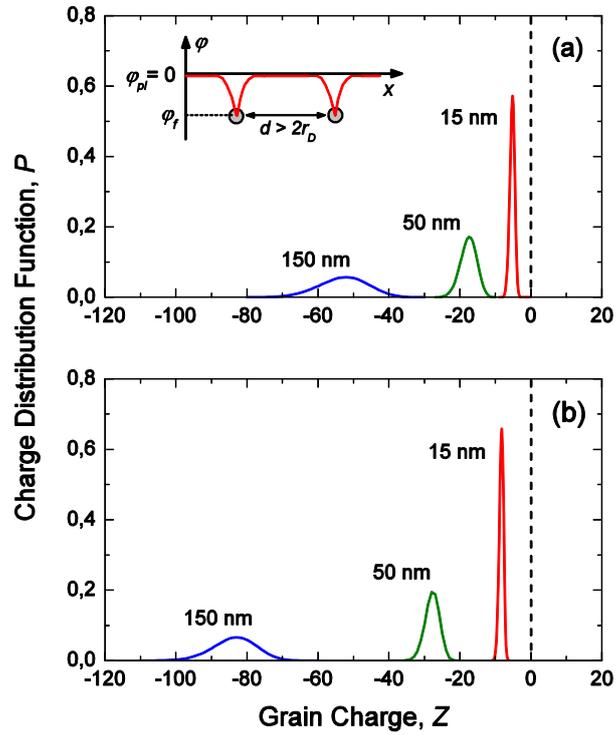

FIG. 1. (Color online) Charge distribution functions in (a) hydrogen and (b) potassium plasma. In both cases $T_e = T_i = 0.2$ eV and $\varphi_{pl} = 0$. The insert shows a schematic diagram of the potential distribution around dust grains in the isothermal plasma.

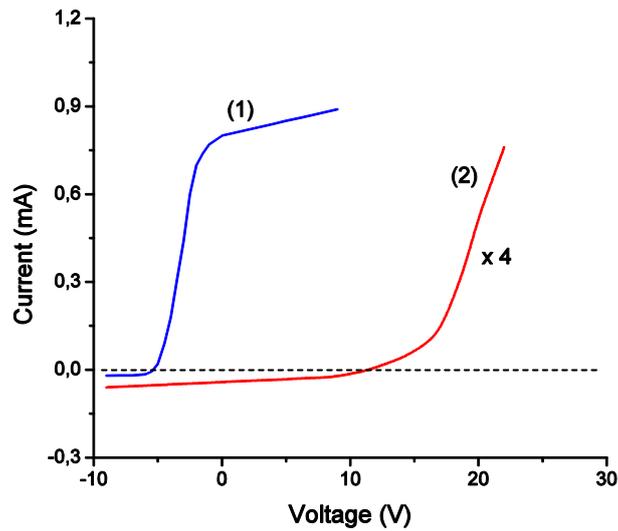

FIG. 2. (Color online) Typical Langmuir probe characteristics (drawn more or less to scale): (1) isothermal ($T_e = T_i = 0.2$ eV) potassium plasma in a Q-machine[26] and (2) non-isothermal ($T_e = 2$ eV $\gg T_i$) argon plasma in an asymmetrical CCP reactor.[31]

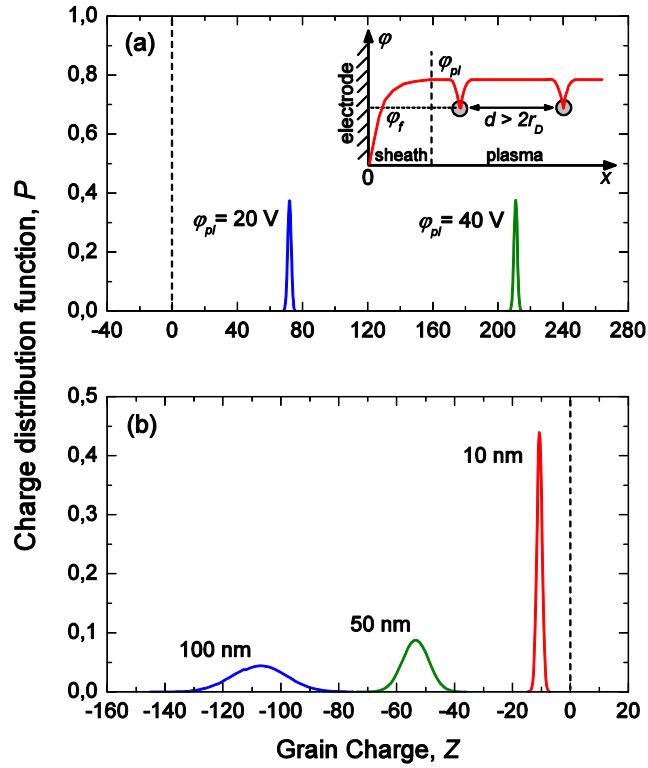

FIG. 3. (Color online) Charge distribution functions in argon plasmas: (a) $a = 10$ nm, $T_e = 4$ eV, $\varphi_{pl} = 20$ and $40$ V) and (b) $T_e = 3$ eV, $\varphi_{pl} = 6$ V, $a = 10, 50$ and $100$ nm. In both cases argon ions are used with $T_i = 0.04$ eV. The insert shows a schematic diagram of the potential distribution near the grounded electrode (or wall) and dust grains in a CCP discharge.

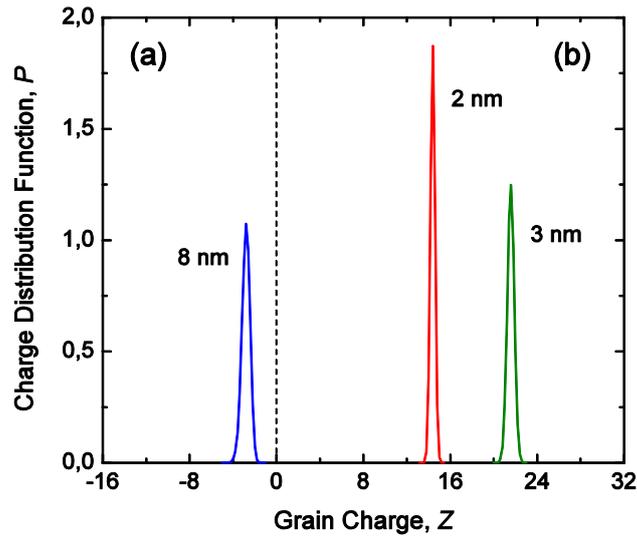

FIG. 4. (Color online) Charge distribution functions of sub-10nm-particles in (a) hydrogen plasma with $T_e = T_i = 0.2$ eV, $\varphi_{pl} = 0$ and (b) argon plasma with $T_e = 4$ eV, $T_i = 0.04$ eV, $\varphi_{pl} = 20$.